\documentclass[a4paper,11pt]{article}
\usepackage{pos}

\newcommand{\vareps}{\varepsilon}

\usepackage{multirow}
\usepackage{tabularx}
\usepackage{mathtools}
\usepackage{booktabs}

\usepackage{physics}
\usepackage{cleveref}
\usepackage{todonotes}

\graphicspath{
	{.} 
	{./figs/}
}

\title{The dependence of observables on action parameters}

\renewcommand{\printHeadAuthors}{G. Catumba}
\author*[a]{Guilherme Catumba}
\author[a]{Alberto Ramos}
\author[a]{Bryan Zaldívar}

\affiliation[a]{Instituto de Física Corpuscular (IFIC) CSIC - Universitat de Valencia. \\ 46071, Valencia, Spain}

\emailAdd{gtelo@ific.uv.es}
\emailAdd{alberto.ramos@ific.uv.es}
\emailAdd{b.zaldivar.m@csic.es}

\abstract{

  Many applications in Lattice field theory require to determine the Taylor
  series of observables with respect to action parameters. A primary example is
  the determination of electromagnetic corrections to hadronic processes. We show
  two possible solutions to this general problem, one based on reweigting, that
  can be considered a generalization of the RM123 method. The other based on the
  ideas of Numerical Stochastic Perturbation Theory (NSPT) in the Hamiltonian
  formulation. We show that 1) the NSPT-based approach shows a much reduced
  variance in the determination of the Taylor coefficients, and 2) That both
  approaches are related by a change of variables. Numerical results are shown for
  the case of $\lambda-\phi^4$ in 4 dimensions, but we expect these observations to be
  general. We conclude by commenting on the possible use of Machine Learning
  techniques to find similar change of variables that can potentially reduce the
  variance in Taylor coefficients.

}

\FullConference{%
The 40th International Symposium on Lattice Field Theory,\\
31st July-4th August, 2023,\\
Fermilab, Batavia, Illinois, USA
}

\begin{document}
\maketitle

\section{Introduction} \label{sec:Introduction}

The use of Monte Carlo (MC) techniques is a useful and common practice in physics, particularly in lattice field theory.
Other than physics, MC methods are broadly used in the machine learning community, in statistical inference, general stochastic optimization problems, among others.
In general, these systems aim at studying an underlying statistical distribution, $p_{\theta}(x)\propto e^{-S(x;\theta)}$, from which we extract observables. On the lattice these involve functional integrals of the form $\expval{\mathcal O} \propto \int dx ~ \mathcal O(x) e^{-S(x;\theta)}$ that are stochastically estimated through MC sampling
\begin{equation}
  \label{eq:expval}
  \expval{\mathcal O(x;\theta)}_{p_{\theta}} = \frac{1}{N}\sum_{\alpha=1}^{N}\mathcal O(x^{\alpha};\theta) + \order{1/\sqrt{N}},
\end{equation}
where the $\{x^{\alpha}\}_{\alpha=1}^{N}$ are samples from the distribution $p_{\theta}(x)$.

As implied by the notation, the distribution functions generally depend on parameters $\theta=\{\theta_{i}\}$ (\textit{e.g.}, bare couplings in case of a lattice field theory). Here we are interested in studying how the observables change under variations of the parameters\footnote{In optimization problems one usually wants to find the optimal value for the parameters under some constraint; in Bayesian inference the sensitivity of the results w.r.t.\ the \textit{hyper-parameters} is important; both cases require the computation of derivatives with respect to the parameters $\theta$ -- see \cite{catumba2023stochastic}}.
The aim of this work is to compute derivatives of the observables with respect to the parameters of the distribution function
\begin{equation}
  \pdv{}{\theta_{i}}\expval{\mathcal O(x;\theta)}_{p_{\theta}},~ \pdv[2]{}{\theta_{i}}{\theta_{j}}\expval{\mathcal O(x;\theta)}_{p_{\theta}},~\dots
\end{equation}
using automatic differentiation techniques.

The available solutions to this problem (see \cite{catumba2023stochastic} for examples) usually require the normalization of the distribution function, preventing its application in conjunction with MC.
Additionally, the use of ordinary Automatic Differentiation \cite{haro} (suited for deterministic functions) is not possible due to the stochastic character of the problem.
In order to extend automatic differentiation to stochastic problems we explore two possible approaches: using reweighting; and by modifying Hamiltonian type sampling algorithms.

\subsection{Automatic Differentiation and Truncated polynomials} \label{sec:Automatic Differentiation and Truncated polynomials}

Conventional automatic differentiation (AD) is a method used to compute derivatives of deterministic functions exactly (up to machine precision).
Forward AD is one of the possible implementations, it exploits the fact that any deterministic function can be built by a finite composition of elementary operations (addition, multiplication, etc.) and elementary functions ($\exp$, $\sin$, etc.).
Defining the derivatives for each fundamental operation and employing the chain rule enables the straightforward computation of any derivative.

In order to implement this idea we start by defining the  order $K$ truncated polynomials, $\tilde x$, as expansions in the variable $\vareps$ of the form
\begin{equation}
\tilde x \equiv x_{0} + x_{1}\vareps + x_{2}\vareps^{2}+\dots+x_{K}\vareps^{K}.
\end{equation}
Secondly, all elementary operations and functions are defined such that all operations are exact at each order:
\begin{align*}
&\tilde x \tilde y = x_{0}y_{0} + (x_{0}y_{1} + x_{1}y_{0})\vareps + (x_{0}y_{2} + 2x_{1}y_{1} + x_{2}y_{0})\vareps^{2} +\dots,\\
&\exp(\tilde x) = e^{x_{0}} + e^{x_{0}}x_{1}\vareps + e^{x_{0}}(x_{1}^{2}/2 + x_{2})\vareps^{2}+\dots,\\
&\dots
\end{align*}
Following this, and by virtue of the Taylor theorem, if we promote a variable $x$ to the truncated polynomial $\tilde x(\varepsilon) = x + \varepsilon$  and evaluate a function $f(x)$ with this variable we get the truncated Taylor series of $f$ at $x$
\begin{align}
  \label{eq:taylor coeffs}
  \tilde f(\varepsilon) = f(\tilde x(\varepsilon)) = \sum_{n}f_{n}\varepsilon^{n}, && f_{n}=\frac{1}{n!}\eval{\pdv[n]{f}{x}}_{x}.
\end{align}

It is important to note that this expansion can be performed in multiple independent variables truncated at different orders -- see \cite{catumba2023stochastic} for a thorough introduction.
The numerical implementation of these rules allows us to extend the ideas of AD to Monte Carlo -- a freely available implementation of the algebra of truncated polynomials used in this work can be found in \cite{alberto_ramos_2023_7970278}.

\subsection{Reweighting \& AD} \label{sec:Reweighting}

The first extension of AD to Monte Carlo combines reweighting techniques with the truncated polynomials.
Starting with samples distributed according to a distribution, $\{x^{\alpha}\} \sim p_{\theta}(x)$, reweighting is a way to compute expectation values w.r.t.\ a different distribution $p_{\theta'}(x)$.
This is done by performing a weighted average
\begin{equation}
  \expval{f(x;\theta)}_{p_{\theta'}} = \frac{\expval{\frac{p_{\theta'}}{p_{\theta}}f(x;\theta)}_{p_{\theta}}}{\expval{\frac{p_{\theta'}}{p_{\theta}}}}
\end{equation}
where the weights $w^{\alpha}\equiv p_{\theta'}(x^{\alpha})/p_{\theta}(x^{\alpha})$ are computed for each sample $\alpha$.
This is an exact relation whose efficiency depends on some measure of similarity between the two distributions.

Since we are interested in computing derivatives w.r.t.\ the parameters $\theta$, $\pdv[n]{}{\theta}$, our method amounts to modifying the target distribution by writing $\theta'$ as a truncated polynomial, $\tilde\theta = \theta + \vareps$.
The new distribution $p_{\theta'}(x) = p_{\tilde\theta}(x)$ becomes a truncated polynomial, carrying the $\theta$-dependence, and consequently the weights $\tilde w^{\alpha}(\vareps)\equiv p_{\tilde\theta}(x^{\alpha})/p_{\theta}(x^{\alpha})$ and the function $f(x;\tilde\theta)$ both become a power series in $\vareps$\footnote{Notice that the coefficients of this power series represent the derivatives w.r.t.\ $\theta$ instead of $x$ as in \cref{eq:taylor coeffs}.}.
When the reweighting average is performed, we obtain a truncated polynomial whose orders give stochastic estimates for the $\theta$-derivatives.

\subsection{Hamiltonian perturbative expansion} \label{sec:Hamiltonian perturbative expansion}

The second method to extend automatic differentiation for MC consists on a modification of Hamiltonian sampling algorithms.
We consider the case of Hybrid Monte Carlo (HMC) \cite{DUANE1987216}, but the implementation is easily mimicked for other algorithms.
The approach is inspired in Numerical Stochastic Perturbation Theory \cite{DIRENZO1994795,brida_smd-based_2017,brida_investigation_2017}, a method used to compute perturbative corrections to a free field theory by solving the molecular dynamics equations at each order in the perturbative coupling.

The HMC algorithm introduces fictitious momenta $\pi$ conjugated to the variables $x$.
Instead of sampling $p_{\theta}(x)\propto \exp{-S(x;\theta)}$, it samples a modified distribution
\begin{align}
  q_{\theta}\propto e^{-H(\pi,x;\theta)}, && H(\pi,x;\theta) = \frac{1}{2}\pi^{2} + S(x;\theta).
\end{align}
Due to the $\pi$-dependence of the Hamiltonian, $H(\pi,x;\theta)$, the momenta are Gaussian distributed.
It is straightforward to see that expectation values that depend only on $x$ remain the same when computed with respect to $q_{\theta}$.
The sampling then follows by repeated iteration of the following sequence: 1) sampling $\pi$ from a Gaussian distribution, $\pi(0)\sim e^{-\pi^{2}/2}$; 2) solving Hamilton's equations of motion (EOM)
\begin{align}
  \dot x = \pdv{H}{\pi} = \pi, &&
  \dot \pi = -\pdv{H}{x},
\end{align}
for a time interval $\tau$ (randomly sampled for ergodicity); and finally 3) accepting the new configuration $(\pi(\tau),x(\tau))$ with probability $e^{-\Delta H}$.
The energy violation $\Delta H = H(\pi(\tau),x(\tau))-H(\pi(0),x(0))$ is due to numerical imprecision when solving the EOM (here we use higher order solvers \cite{OMELYAN2003272}).
The chain of values $x(\tau)$ obtained in this way are distributed according to $p_{\theta}(x)$.

The extension for automatic differentiation starts by 1) promoting the parameters to truncated polynomials, $\tilde\theta(\vareps) = \theta +\vareps$.
This implies that the remaining variables also become expansions, $x,~\pi\rightarrow \tilde x(\vareps),~\tilde \pi(\vareps)$.
Momentum refresh is done at zeroth order only, $\tilde\pi(0)\sim\exp(\pi^{2}(0)/2)$, while the remaining orders are initialized to zero.
2) The equations of motion are solved at each order (this is straightforward when using a numerical implementation of the algebra of truncated polynomials);
3) The energy violation $\Delta H$ is a truncated polynomial, not allowing to perform the accept/reject step.
The results require an extrapolation to zero integration step size, however, in practice it is  enough to work at a sufficiently precise step size such that the possible extrapolation is bellow the statistical error.

With this modification, we obtain samples $\{\tilde x^{\alpha}\}_{\alpha=1}^{N}$ which are truncated polynomials, carrying the $\theta$ dependence.
Ordinary expectation values correspond to a Taylor expansion around $\theta$
\begin{equation}
  \frac{1}{N}\sum_{\alpha=1}^{N}f(\tilde x^{\alpha};\tilde\theta) = \expval{f(x;\theta)} + \pdv{}{\theta}\expval{f(x;\theta)} \vareps+\dots
\end{equation}
such that we recover the corresponding derivatives.

\section{Example in lattice 4D scalar theory} \label{sec:Example in 4D lattice scalar theory}

We consider a $\lambda-\phi^{4}$ lattice theory in 4 space-time dimensions to test both methods.
The lattice Euclidean action is
\begin{equation}
  S_{\rm latt}(\phi; m,\lambda) = \sum_x \left\{  \frac{1}{2}\sum_\mu[\phi(x+\mu) - \phi(x)]^2 + \frac{ m^2}{2}\phi^2(x) + \lambda\phi^4(x) \right\}
\end{equation}
with the field $\phi$ and mass $m$ being dimensionless quantities.

The simulations\footnote{The code used to generate the data can be found at \url{https://igit.ific.uv.es/alramos/lambdaphi4.jl/}} are perfomed on a $L^{4}$ lattice with $L/a=32$.
This is done for several values of $\lambda$ at fixed $m^{2}=0.05$.
We are interested in estimating the dependence of expectation values of this theory on the couplings $m^{2}$ and $\lambda$.
For this we promote $m^{2}\rightarrow \tilde m^{2}$ and $\lambda\rightarrow \tilde \lambda$ (a double expansion around $m^{2}$ and $\lambda$) and perform reweighting from samples computed at a fixed value of the couplings.
On the other hand, we also generate samples, $\{\tilde\phi^{\alpha}(x)\}_{\alpha=1}^{N}$, from the Hamiltonian expansion, where $\tilde\phi$ is a truncated polynomial in $m^{2}$ and $\lambda$ (a single HMC simulation allows to compute both derivatives).

As observables, we consider expectation values of the field and the action density
\begin{align}
\expval{\phi^{2}(x)} && \langle s(x;m,\lambda) \rangle =  \frac{1}{2}\langle [\phi(x+\mu) - \phi(x)]^2\rangle + \frac{ m^2}{2}\langle \phi^2(x)\rangle + \lambda\langle \phi^4(x)  \rangle\,.
\end{align}
Notice that in this way we test the method for observables with and without `connected' contributions, \textit{i.e.}, contributions for the derivative coming from the explicit dependence of the observables w.r.t.\ the parameters $m^{2}$ or $\lambda$.
In \cref{tab:lp4} we show the results for the first derivatives, $\partial_{m^{2}},~\partial_{\lambda}$, and cross derivative $\partial^{2}_{m^{2},\lambda}$ computed with the reweighting and Hamiltonian approaches at various values of $\lambda$ and for fixed $m^{2}=0.05$.

\begin{table}
  \centering
    \scalebox{0.8}{
  \begin{tabular}{llllllll}
    &&&\multicolumn{5}{c}{$\lambda$} \\
    \cline{4-8}
    &&&0.0&0.1&0.2&0.3&0.4\\
    \midrule
    \multirow{6}*{$\langle \phi^2 \rangle$}& \multirow{2}*{$\partial_{m^2}$}
     & RW &     -0.0428(20)&     -0.0328(14)&     -0.0270(13)&     -0.0241(12)&     -0.0220(11) \\
    && HAD &   -0.042526(41)&   -0.030880(14)&   -0.026273(10)&  -0.0233672(82)&  -0.0212721(72)\\
    \cline{2-8}
    &\multirow{2}*{$\partial_{\lambda}$}
     & RW &     -0.0779(22)&    -0.05227(94)&    -0.04370(89)&    -0.03534(61)&    -0.03169(50)\\
    && HAD &   -0.077816(79)&   -0.052499(24)&   -0.042218(19)&   -0.035830(14)&   -0.031323(11)\\
    \cline{2-8}
    &\multirow{2}*{$\partial^2_{m^2,\lambda}$}
     & RW &        0.43(43)&        0.03(16)&        0.16(14)&       -0.10(11)&       0.116(77)\\
    && HAD &      0.2733(22)&    0.061593(99)&    0.035082(69)&    0.024240(42)&    0.018263(31)\\
    \midrule
    \multirow{6}*{$\langle s \rangle$}&\multirow{2}*{$\partial_{m^2}$}
     & RW & -0.0025(42)&           -0.0006(34)&       0.0027(35)&       0.0028(36)&       0.0057(32) \\
    && HAD &    -0.000003(22)&     0.002623(16)&     0.004218(20)&     0.005397(14)&     0.006265(16) \\
    \cline{2-8}
    &\multirow{2}*{$\partial_{\lambda}$}
     & RW &    -0.0765(46)&     -0.0567(26)&     -0.0538(25)&     -0.0447(23)&     -0.0400(17) \\
    && HAD &   -0.069774(49)&   -0.057738(34)&   -0.050128(40)&   -0.044530(27)&   -0.040250(27) \\
    \cline{2-8}
    &\multirow{2}*{$\partial^2_{m^2,\lambda}$}
     & RW &     -1.9(2.8)&        -0.16(39)&         0.36(43)&        -0.43(32)&         0.16(26) \\
    && HAD &     0.038864(96)&     0.019197(66)&     0.013405(69)&     0.010126(50)&     0.007860(47) \\
    \bottomrule
  \end{tabular}
  }
  \caption{First derivatives with respect to $m^2$ ($\partial_{m^2}$), $\lambda$ ($\partial_\lambda$) and the cross derivative ($\partial^2_{m^2,\lambda}$) for $\expval{\phi^{2}}$ and $\expval{s}$ from a $L^4$ lattice with $L/a=32$ and $m^2 = 0.05$. The reweighting method is denoted RW while HAD corresponds to the Hamiltonian expansion.}
  \label{tab:lp4}
\end{table}

The first remarkable thing to notice is that the Hamiltonian approach delivers more precise results across all available parameter space.
In fact, although both methods were obtained with the same statistics, compared to the reweighting method the Hamiltonian expansion shows a factor of $\sim100$ in precision for the derivatives.
The $m^{2}$-derivative of the action density, which should be zero for $\lambda=0$, is a good example of the level of precision of both methods.
The situation is even more severe for the cross derivatives where the reweighting approach is not able to deliver a consistent signal while the Hamiltonian method maintains some precision.

It is important to notice that both methods are valid for derivatives of arbitrary orders.
In the original work \cite{catumba2023stochastic} the truncated Taylor expansion was shown to properly reconstruct the parameter dependence for a significant range, which is dictated by the underlying function, the truncation order, and the available statistics.

\subsection{Comparison of methods} \label{sec:Method comparison}

What is the reason for the difference in the precision of both approaches?
To understand this it is worth summarizing their properties and attempt to build a connection between the methods.
We start with the reweighting for an observable $\mathcal O$ with weights $w=e^{-\delta S}$, whose formula at leading order is
\begin{equation}
      \partial_{\theta} \expval{\mathcal O} = \expval{\partial_{\theta} \mathcal O} + \left[ \expval{\mathcal O}\expval{\partial_{\theta} S}- \expval{\mathcal O\partial_{\theta} S} \right].
\end{equation}
The first term is the connected part, while the one in parenthesis corresponds to the disconnected contribution.
The latter comes from the implicit dependence of the distribution function on the parameters $\theta$ and usually suffers from large variance leading to less precise results.
While the reweighting method re-utilizes conventional samples from $p_{\theta}(x)$, the Hamiltonian samples, $\{\tilde\phi\}$, carry their dependence on the parameters $\theta=\{m^{2},~\lambda\}$ in the form of a Taylor series, thereby avoiding the computation of disconnected terms.

\begin{figure}[htb]
\centering
\includegraphics[scale=0.80]{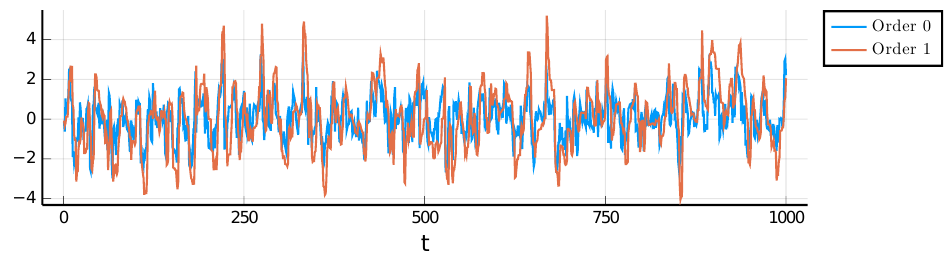}
\caption{Solution of the equations \cref{eq:eomorders0,eq:eomorders1} with a trajectory length tuned to reduce the variance and autocorrelations.}
\label{fig:eom}
\end{figure}

Another way to see this and connect both methods uses the reparametrization trick.
Consider the following toy model, a Gaussian distribution function, $p_{\sigma}(x) = \frac{1}{\sigma\sqrt{2\pi}} e^{-\frac{x^2}{2\sigma^2}}$ where $\theta=\sigma$.
Having samples $\{x^{\alpha}\}\sim p_{\sigma^{*}=1}$ we can obtain samples from another value of $\sigma$ by a change of variables, $y^{\alpha}=\sigma x^{\alpha}$, $\{y^{\alpha}\}\sim p_{\sigma}$.
If we apply reweighting with AD on the $\{x^{\alpha}\}$ samples, using an expansion around $\sigma^{*}=1$ as $\tilde\sigma= \sigma^{*}+\vareps$, the weights become
\begin{equation}
  \tilde w(\vareps) \propto \exp(-\frac{x^{2}}{2}\left(\frac{1}{(\sigma^{*}+\vareps)^{2}} - \frac{1}{(\sigma^{*})^{2}}\right)).
\end{equation}
However, if we apply reweighting, $\sigma^{*}=1 \rightarrow \sigma$, after changing to the $y$-variables, the reweighting factors become trivially $y$-independent
\begin{equation}
  \tilde w(\vareps) = \frac{p_{\tilde\sigma}(y)}{p_{\sigma^{*}}(y)} = \frac{1}{\sigma^{*}+\vareps}
\end{equation}
thus being irrelevant in the computation of the expectation values.
The fact that there is no effective reweighting is clear since we use the correctly sampled variables.
To understand the connection with the Hamiltonian expansion, we start by writing the EOM for the first two orders
\begin{align}
  \label{eq:eomorders0}
&\ddot x_0 = - \frac{x_0}{\sigma^2},\\
  \label{eq:eomorders1}
&\ddot x_1 = - \frac{x_1}{\sigma^2} + 2\frac{x_0}{\sigma^3}
\end{align}
(here we suppressed the momentum variables in favour of a second order differential equation).
While the zeroth order is an harmonic oscillator, the leading order equation corresponds to an undamped driven oscillator whose driving force has the same frequency as the oscillator\footnote{This also leads to a resonant phenomena, where the oscillation amplitude grows with the integration time. Although for any finite integration time it does not affect the central value of the expectation value, it may lead to large variance. The average trajectory length was optimized to reduce the variance and autocorrelations.} -- this leads to both orders having the same average value (over various trajectories), $\expval{x_{1}} = \expval{x_{0}}$.
This can be seen in \cref{fig:eom} where the solutions of the EOM for the first two orders are shown to be similar.

The key to understand the relation between reweighting, Hamiltonian expansion, and the change of variables lies in the truncated polynomial corresponding to $y^{\alpha}$ from the reparametrization, $y^{\alpha} =\sigma x^{\alpha} \longrightarrow \tilde y^{\alpha} = x^{\alpha} + x^{\alpha} \vareps$.
Notice how the first two orders are the same, exactly as in the solutions of the EOM above.
We conclude that the EOM from the Hamiltonian approach find the transformation/change of variables that leads to constant reweighting factors, and thus the variance of the method is not affected by the disconnected contributions arising from the weighting factors.

\subsection{Improved reweighting} \label{sec:Improved reweighting}

Given that the conventional reweighting suffers from large variance, and knowing that the variance for the derivatives is dependent on the change of variables, we may attempt to improve the reweighting method by finding a suitable transformation.
A similar approach, using a gradient flow transformation for QCD can be found in \cite{bacchio2023novel}.

Let us again consider the $m^{2}$-derivatives for the scalar theory.
The $m^{2} \rightarrow \tilde m^{2}$ reweighting factors are
\begin{equation}
    \tilde w^{\alpha} = \exp{- S(x^{\alpha}; \tilde m^{2},\lambda) + S(x^{\alpha}; m^{2},\lambda)}
\end{equation}
(here we consider only an expansion in $m^{2}$).
If we first change variables to $\tilde y = x + f(x)\vareps$ with a unknown function $f$, the weights become
\begin{equation}
    \tilde w^{\alpha} =  \exp{S(x^{\alpha}; m^{2},\lambda) - S(\tilde y^{\alpha}; \tilde m^{2},\lambda) - \log\abs{\tilde J}},
\end{equation}
where the last term comes from the Jacobian of the transformation.
The aim is to find the a suitable transformation $f$ to reduce the effect of reweighting.

We can find this transformation exactly for the free theory, whose lattice action in momentum space reads
\begin{align}
    S_{\textrm{latt}}(\phi_{p}; m) = \sum_{p} \phi_{p}^{*}\left[ \sum_{\mu}\hat p_{\mu}^{2} +  m^{2} \right]\phi_{p}, && \hat p =2\sin(ap/2).
\end{align}
The change of variables in momentum space becomes $\tilde\varphi_{p} = \phi_{p} +f_{p}\vareps$.
The term $\Delta S$ in the reweighting factor
\begin{equation}
  S(\phi_{p}; m^{2}) - S(\tilde \varphi_{p}; \tilde m^{2}) = -\vareps\left[  (\hat p^{2} + \tilde m^{2})(\phi_{p}^{*}f_{p}+ \phi_{p}f_{p}^{*}) + \phi_{p}^{*}\phi_{p} \right]
\end{equation}
can be eliminated with the $f_{p} = - \frac{\phi_{p}}{2(\hat p^{2}+m^{2})}$.
With this choice the weights have the Jacobian contribution only which turns out to be $\phi$-independent $w(\phi_{p};m) = \prod_{p}- \frac{1}{2(\hat p^{2}+m^{2})}$,
and thus not affecting the reweighting.

\begin{table}
  \centering
    \scalebox{0.8}{
  \begin{tabular}{llllllll}
    &&&\multicolumn{5}{c}{$\lambda$} \\
    \cline{4-8}
    &&&0.0&0.1&0.2&0.3&0.4\\
    \midrule
    \multirow{3}*{$\langle \phi^2 \rangle$}&\multirow{3}*{$\partial_{\hat m^2}$}
     & RW &      -0.0428(20)&      -0.0328(14)&      -0.0270(13)&      -0.0241(12)&      -0.0220(11) \\
    && TRW &      -0.042528(30) &      -0.03069(32) &      -0.02604(37) &      -0.02328(54) &      -0.02079(56)\\
    && HAD &    -0.042526(41)&    -0.030880(14)&    -0.026273(10)&   -0.0233672(82)&   -0.0212721(72)\\

    \midrule
    \multirow{3}*{$\langle s \rangle$}&\multirow{3}*{$\partial_{\hat m^2}$}
     & RW &          -0.0025(42)&           -0.0006(34)&       0.0027(35)&       0.0028(36)&       0.0057(32) \\
    && TRW  &               3(28)$\times 10^{-19}$ &        0.00344(85) &         0.0052(11) &        0.0052(17) &        0.0072(15) \\
    && HAD &    -0.000003(22)&     0.002623(16)&     0.004218(20)&     0.005397(14)&     0.006265(16) \\
    \bottomrule
  \end{tabular}
  }
  \caption{Derivatives with respect to $\hat m^2$ for $\expval{\phi^{2}}$ and $\expval{s}$ from a $L^4$ lattice with $L/a=32$ and $\hat m^2 = 0.05$. The derivatives were computed with conventional reweighting, transformed reweighting (TRW), and with the Hamiltonian expansion (HAD).}
  \label{tab:lp4_trw}
\end{table}

In \cref{tab:lp4_trw} the $m^{2}$-derivative from reweighting after the transformation is compared with the normal reweighting and the Hamiltonian approach.
For the free theory we obtain the best results since the transformation is exact.
In fact, for $\expval{\phi^{2}}$ the precision of the Hamiltonian expansion is similar to the improved reweighting, which supports the idea that the former method finds the correct change of variables.
Nonetheless, although the transformation is not exact for non-zero $\lambda$ values, it still constitutes an improvement in relation with conventional reweighting for all available values of the coupling, with a decrease in precision for the largest values.

\section{Conclusion} \label{sec:conclusion}

While automatic differentiation is a very important tool across many areas, there are no simple extensions when Monte Carlo sampling is involved in the definition of the function.
These problems involve an underlying distribution function that depends on some parameters (couplings in the context of lattice field theory).
The dependence on these parameters is very useful and usually difficult to determine.
In this work we suggest two possibilities that tackle the problem by using truncated polynomials.

One of the methods is a straightforward modification of reweighting averages with the implementation of truncated polynomials.
It has the advantage of being agnostic of the sampling algorithm, but may in general lead to large variance problems which are in general hard to predict.

The second method consists on a modification of Hamiltonian sampling algorithms. In this case we produce samples that carry the dependence on the parameters in the form of Taylor expansions.
We showed (for HMC) that this method is able to reach a much higher precision with the same level of statistics.
It is important to refer that the convergence of the equations of motion is model dependent.
For the example considered in this work the convergence is guaranteed.

Both methods were tested on a 4-dimensional lattice scalar theory, where the dependence on the mass and quartic coupling was studied.
Additionally, we have proposed that the Hamiltonian approach can be seen as a transformation to a basis where the reweighting factors are constant.
This sustains the claim that all parameter dependence is in the samples and that the absence of disconnected contributions is the reason behind lower variance in the second method.

This connection allowed us to improve the reweighting method by first transforming the variables.
Although finding these transformations is, in general, a highly non-trivial task, this is a suitable problem for the application of Machine-Learning techniques, which can significantly improve the computation of derivatives of observables.

\section*{Acknowledgements}

AR and GT acknowledge financial support from the Generalitat
Valenciana (CIDEGENT/2019/040). Similarly, BZ akcnowledges the support from CIDEGENT/2020/055.
The authors gratefully acknowledge as well the support from the Ministerio de
Ciencia e Innovacion (PID2020-113644GB-I00) and computer resources at
Artemisa, funded by the European Union ERDF and Comunitat Valenciana
as well as the technical support provided by the Instituto de Física
Corpuscular, IFIC (CSIC-UV). The authors acknowledge the financial support from the MCIN with funding from the European Union NextGenerationEU (PRTR-C17.I01) and Generalitat Valenciana. Project ``ARTEMISA'', ref. ASFAE/2022/024.

\end{document}